\documentclass[conference]{IEEEtran}
\IEEEoverridecommandlockouts
\usepackage{cite}
\usepackage{amsmath,amssymb,amsfonts}
\usepackage{algorithmic}
\usepackage{graphicx}
\usepackage{textcomp}
\usepackage{xcolor}
\def\BibTeX{{\rm B\kern-.05em{\sc i\kern-.025em b}\kern-.08em
    T\kern-.1667em\lower.7ex\hbox{E}\kern-.125emX}}
\begin{document}

\title{Anomaly Detection in Particle Accelerators using Autoencoders} 

\author{\IEEEauthorblockN{1\textsuperscript{st} Jonathan P. Edelen}
\IEEEauthorblockA{
\textit{RadiaSoft LLC}\\
Boulder, USA\\
jedelen@radiasoft.net}
\and
\IEEEauthorblockN{2\textsuperscript{nd} Nathan M. Cook}
\IEEEauthorblockA{ 
\textit{RadiaSoft LLC}\\
Boulder, USA \\
ncook@radiasoft.net}
}

\maketitle

\begin{abstract}
The application of machine learning techniques for anomaly detection in particle accelerators has gained popularity in recent years. These efforts have ranged from the analysis of quenches in radio frequency cavities and superconducting magnets to anomalous beam position monitors, and even losses in rings. Using machine learning for anomaly detection can be challenging owing to the inherent imbalance in the amount of data collected during normal operations as compared to during faults. Additionally, the data are not always labeled and therefore supervised learning is not possible. Autoencoders, neural networks that form a compressed representation and reconstruction of the input data, are a useful tool for such situations. Here we explore the use of autoencoder reconstruction analysis for the prediction of magnet faults in the Advanced Photon Source (APS) storage ring at Argonne National Laboratory.

\end{abstract}


\section{Introduction}

In recent years machine learning (ML) has been identified as having the potential for significant impact on the modeling, operation, and control of particle accelerators~\cite{edelen_etal:2016, Edelen:2018jid}. These techniques are attractive due to their ability to model nonlinear behavior, interpolate on complicated surfaces, and adapt to system changes over time. This has led to a number of dedicated efforts to apply ML and early efforts have shown promise. 

For example, neural networks have been used as surrogates for traditional accelerator diagnostics to generate non-interceptive predictions of beam parameters\cite{PhysRevAccelBeams.21.112802, edelen_etal:2016b}. Neural networks have been used for a range of machine tuning problems utilizing inverse models \cite{edelen_etal:2017, edelen_etal:2019}. When used in conjunction with optimization algorithms these have demonstrated improved switching times between operational configurations \cite{PhysRevLett.121.044801}. Neural networks have also been demonstrated to significantly speed up multi-objective optimization of accelerators by using them as surrogate models \cite{PhysRevAccelBeams.23.044601}. 

Anomaly detection has also been specifically highlighted as an area where machine learning can significantly impact operational accelerators. These algorithms work by identifying subtle behaviors of key variables prior to negative events. There have been many efforts to apply ML tools for anomaly detection across accelerators and accelerator subsystems. For example, understanding and predicting faulty behavior in superconducting radio frequency (RF) cavities and magnets is of interest due to the potential catastrophic nature of a failure of these devices. ML tools have been applied to detect anomalies in superconducting magnets at CERN \cite{WIELGOSZ201740} and RF cavities at DESY \cite{7739750, NAWAZ20181379, Nawaz:2018hms}. Additionally, machine learning has been used to identify and remove malfunctioning beam position monitors in the Large Hadron Collider (LHC), prior to application of standard optics correction algorithms \cite{Fol:2693725}. Other efforts have sought to use ML for detection of errors in hardware installation \cite{DewitteThiebout2019ADfC}. 

While many of these efforts have shown success, results for global fault prediction have been limited. A recent effort at J-PARC utilized the System Invariant Analysis Technique to develop an operational fault prediction algorithm \cite{Soma2017}. However, these results are preliminary, and loss classification and fault prediction remain active areas of research. Loss prediction has been studied at the LHC \cite{Valentino_2017} but the use of autoencoders for fault prediction has yet to be fully explored. In this paper we evaluate the use of autoencoders to identify precursors to magnet failures in the Advanced Photon Source (APS) storage ring. We begin with an overview of autoencoders and our methodologies followed by a discussion of the APS accelerator chain and our dataset. We then build autoencoder models and demonstrate their efficacy using both unsupervised learning and semisupervised learning. 

\section{Autoencoders} 

\begin{figure}[htbp]
\centerline{\includegraphics[width = \columnwidth]{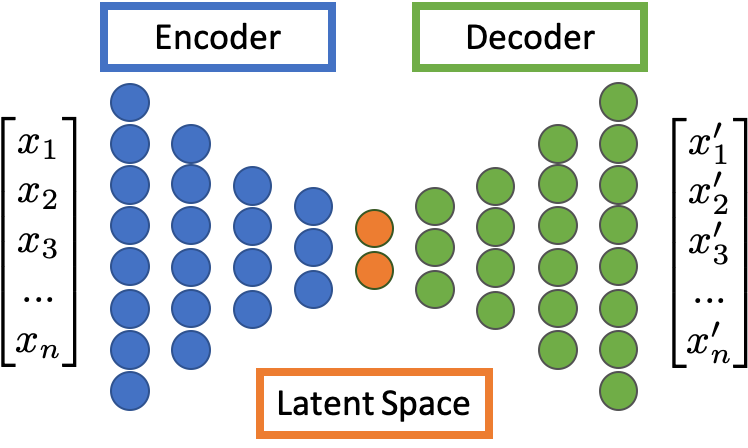}}
\caption{Schematic digram of an autoencoder. The inputs noted by $x_n$ and the outputs also noted as $x'_n$. The encoder section is highlighted in blue, the latent space in orange, and the decoder in green. In this study, we consider fully connected layers, but omit the connections to enhance figure clarity.}
\label{fig:1}
\end{figure}

Autoencoders are a class of neural networks that seek to reconstruct an input dataset while simultaneously reducing its dimensionality. The two main characteristics of the autoencoder are that the inputs are equal to the outputs, and the waist of the network is smaller in dimension than the input dataset. Figure \ref{fig:1} shows a schematic of an autoencoder. Here the number of nodes is steadily decreased in the encoder section (Blue). The encoded dimension (Orange), also referred to as the latent space, is the minimum number of nodes. The number of nodes per layer then increases in the decoder section (Green) to reproduce the input data. The base dimensionality of the dataset is determined by the number of nodes at the encoded dimension. Typically the encode and decode sections of the network are symmetric. While many types of autoencoders can be constructed using feed-forward layers, convolutional layers, recurrent layers, or a combination of layers, the focus of this paper is on vanilla feed forward neural networks. 

Autoencoders are commonly used in two configurations. The first is the direct analysis of the latent space. This is accomplished by removing the decoder section from the network and analyzing the output of the latent space nodes directly. The second configuration is used to quantify the relationship between a training dataset and a test dataset. Here one evaluates the ability for the autoencoder to reconstruct a given input data set. This provides a quantifiable metric for how similar a new dataset is to the training data either with respect to individual input parameters or in aggregate by computing the root mean squared (RMS) reconstruction error, $\sqrt{ 1 / n \Sigma_n (x_i - x'_i)^2}$.

\section{The Advanced Photon Source}  
\begin{figure*}[htbp]
\centerline{\includegraphics[width = \textwidth]{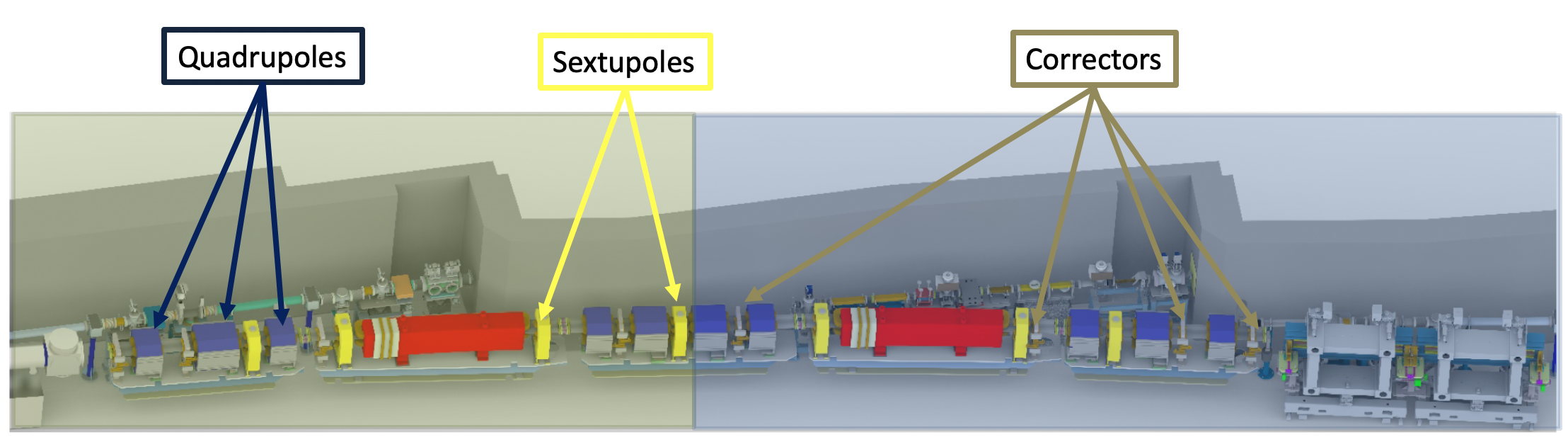}}
\caption{Schematic of the APS storage ring. Here we show the A and B sectors for one of the larger numbered sectors. The delineation between A and B sectors is noted by the yellow and blue shading. The magnet types are color coded with quadrupoles in Blue, sextupoles in yellow and correctors in brown. The dipole magnets noted in red are not included in our study.}
\label{fig:3}
\end{figure*}

The Advanced Photon Source consists of an electron accelerator chain that produces a bunched beam at 7 GeV energy, which traverses a periodic storage ring to generate focused radiation for light source end users. The main storage ring contains over one thousand different magnets that are all individually powered. The primary goal of these magnets is to maintain the optical properties of the beam in order to ensure proper delivery of photons to users. The ring is broken down into forty numbered sectors each with an A and B sub-sectors resulting in a total of 80 individual sectors. Figure \ref{fig:3} shows a schematic diagram of two sectors in the storage ring. 

The data used for our study were curated over a three year period of running. The data are broken down into two categories, the reference data and the fault data. For the reference data the storage ring was operating under normal conditions, without any faults or beam loss. The fault data contains information on the magnets leading up to a magnet fault. Here, one of the magnets in the ring fails resulting in beam loss and an unexpected down time. For the purposes of our analysis the actual seconds leading up to and including the fault are excluded. This allows us to verify the identification of precursors that can be used for fault prediction in the future. For each magnet there are current, voltage, and temperature measurements. These measurements are logged at 1Hz. 

\begin{figure}[h!]
\centerline{\includegraphics[width = \columnwidth]{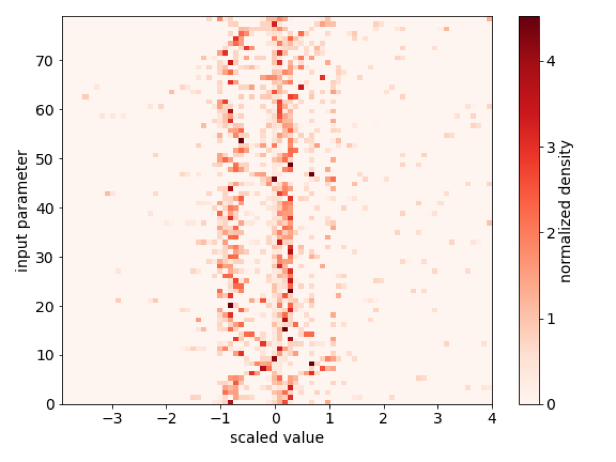}}
\caption{Heat map of the reference data used to train the autoencoder. The vertical axis denotes the sector number and the horizontal axis denotes the sample magnitude. The color scale indicates the frequency of those magnitudes aggregated over the entire run period. }
\label{fig:4a}
\end{figure}

Because there are a large number of magnets and each magnet has multiple diagnostics we simplified the input dataset in order to narrow the focus of our initial investigation. For each sector we computed the sum of the currents over all magnets in the sector at each time-step. This results in 80 input parameters corresponding to the effective magnet current for each sector. Figure \ref{fig:4a} shows a heat map of the magnet current values for each of the 80 sectors for the reference dataset. 

\begin{figure}[h!]
\centerline{\includegraphics[width = \columnwidth]{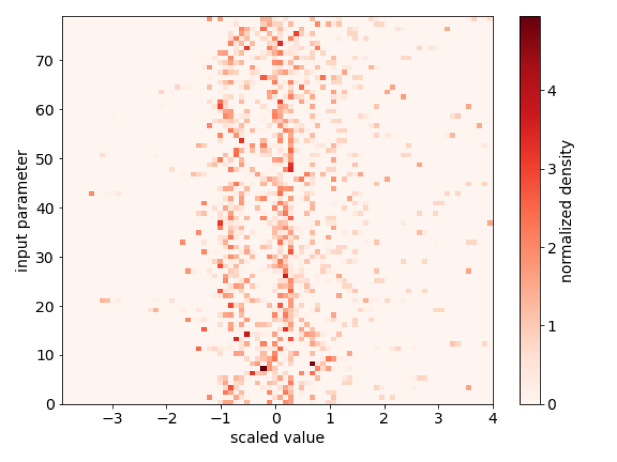}}
\caption{Heat map of the test data used to train the autoencoder. The vertical axis denotes the sector number and the horizontal axis denotes the sample magnitude. The color scale indicates the frequency of those magnitudes aggregated over the entire run period.}
\label{fig:4b}
\end{figure}

Figure \ref{fig:4b} shows a heat map of the magnet current values for each of the 80 sectors for the test dataset. Visual comparison of Figures \ref{fig:4a} and \ref{fig:4b} show clear differences in the datasets that should be well characterized by the autoencoder. The bulk structure for these two datasets is similar and the variation from sector to sector is small. This indicates that while the autoencoder will likely be able to identify anomalies in aggregate, differentiating between sector-specific anomalies will be more challenging. For our initial analysis, we only consider aggregate detection, and consider individual sector analysis in Section~\ref{sec:fault}.

\section{Model Learning and Evaluation} 

\begin{figure}[htbp]
\centerline{\includegraphics[width = \columnwidth]{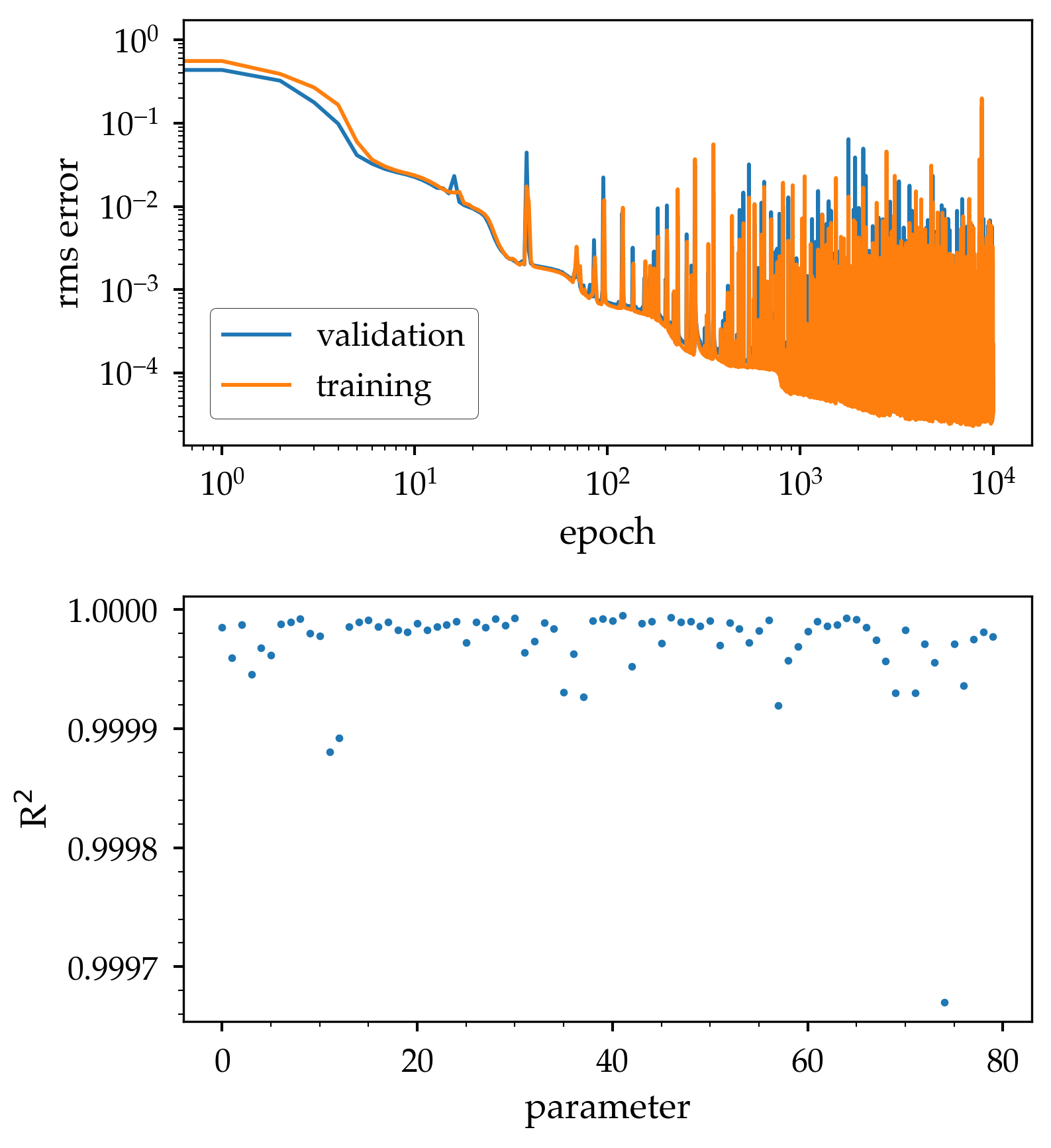}}
\caption{Top: Training and validation loss for the autoencoder as a function of epoch. The network is trained in mini-batches where the batch size was 1000. Bottom: $R^2$ value as a function of input parameter for the validation set. For a perfectly trained model the relationship between the predicted value and the input value should be linear giving an $R^2$ of 1. Here we see very good agreement between the input data and the output data for the autoencoder. }
\label{fig:5}
\end{figure}

The autoencoder was trained and validated on the reference data using an 80/20 split. The fault data was split into two test sets using roughly a 50/50 split. While the training and validation split was conducted randomly, the test set was split by run ensure robustness in the semisupervised learning studies. Because our dataset is not uniformly sampled we utilized a robust scalar to scale the data prior to training, validation, and testing. The robust scaler removes the median value and scales the data by the interquartile range. Note that the scaler is fit to the reference data and applied to the test data before sending it to the model. Due to the fact that the inputs and outputs are not between -1 and 1 rectified linear units will be the most appropriate choice of activation function. The network architecture consisted of 5 layers in the encoder starting from 60 nodes and decreasing to 10 nodes in increments of 10. The latent space dimension was 5 nodes and the decoder section was a mirror of the encoder section. Figure \ref{fig:5} shows the training and validation loss as a function of training epoch (top) and the $R^2$ value for the reconstruction of each parameter (bottom). 

\begin{figure}[htbp]
\centerline{\includegraphics[width = \columnwidth]{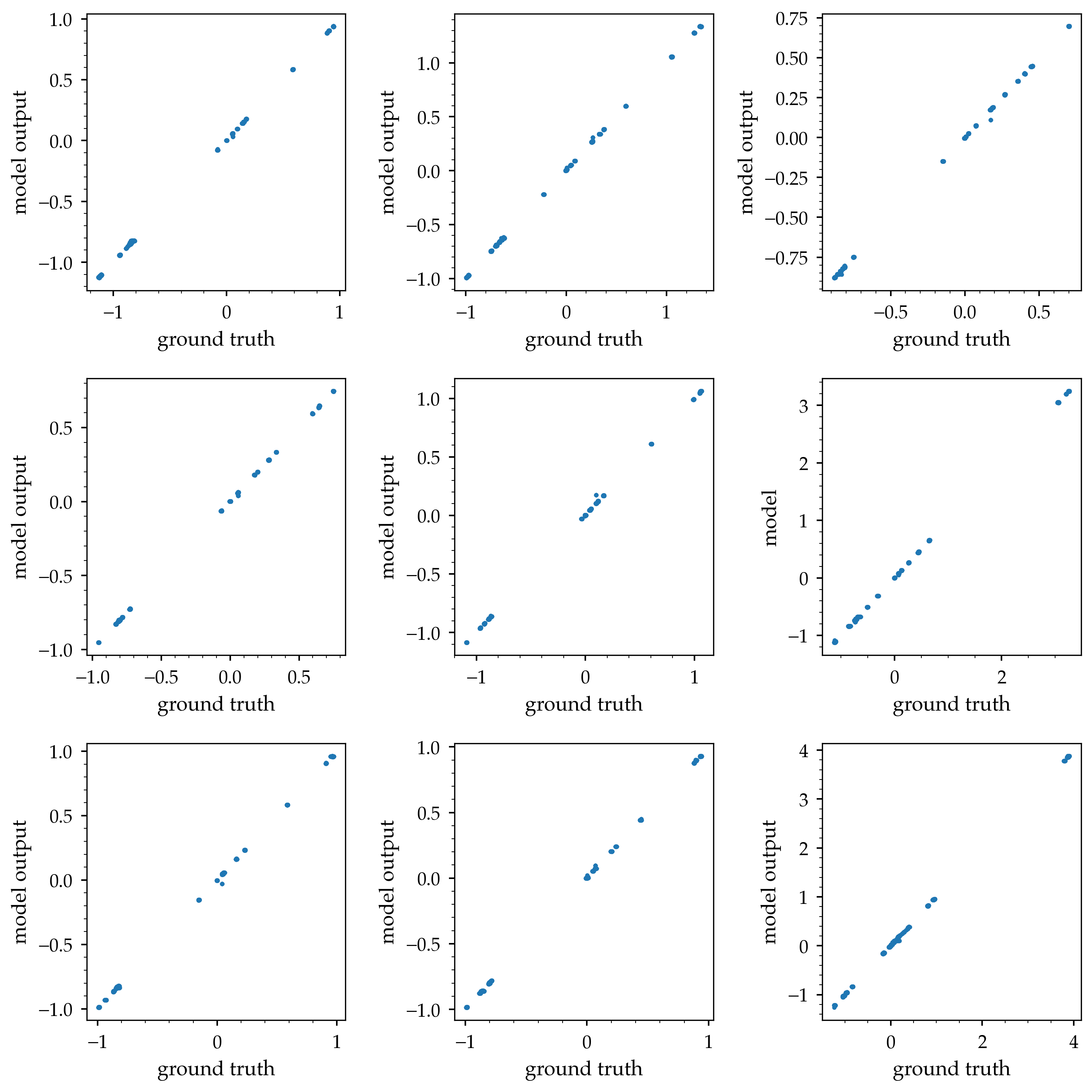}}
\caption{Model prediction as a function of the ground truth for nine randomly chosen sectors.}
\label{fig:5a}
\end{figure}

Figure \ref{fig:5a} shows the predicted input parameter against the ground truth input parameter for nine of the sectors using the validation data. The linear relationship between the model output and the input ground truth here show that our autoencoder is well trained. 

\begin{figure}[htbp]
\centerline{\includegraphics[width = \columnwidth]{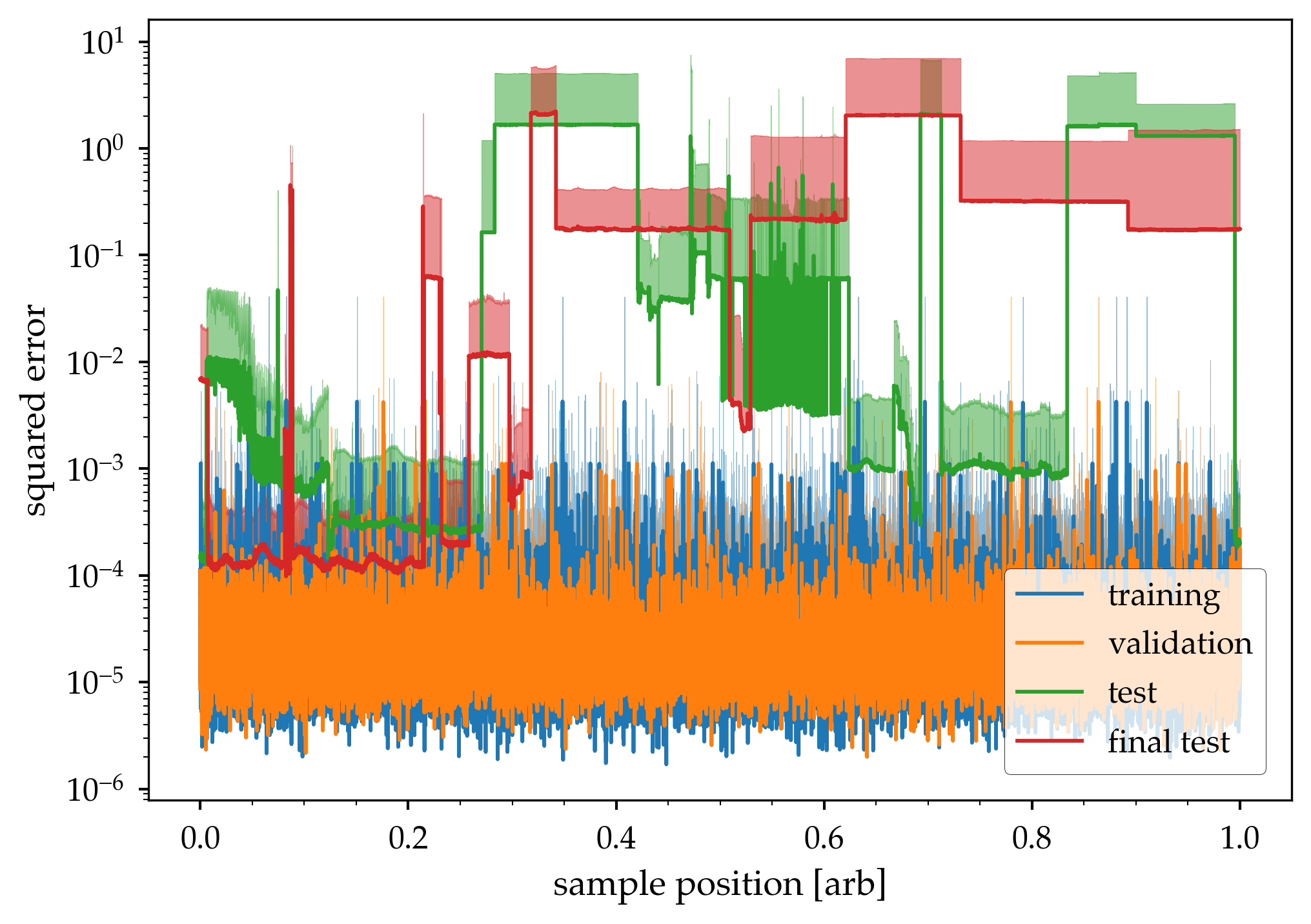}}
\caption{Mean squared error across all sectors as a function of sample position for the reference data (training and validation) and fault data (test and final test). The green dataset is used later for tuning detection thresholds in the semisupervised case while the red data are held aside for final testing. The shaded region indicates the variance in the squared error across the 80 sectors. Because the squared error is always positive the variance is only shown in one direction.} 
\label{fig:6}
\end{figure}

In order to identify fault precursors we computed the reconstruction error in two ways. The first uses the squared error of each sector and the second uses the root mean squared error (RMS) over all 80 sectors. We begin with a comparison of the two different error calculations on the reference data and the fault data. Figure \ref{fig:6} shows the squared reconstruction error as a function of sample position for the reference data and the fault data. The shaded region depicts the variance in the squared reconstruction error across all 80 sectors. Note that the runs are concatenated in time and the samples span many different runs over the aforementioned three year period. The sample positions are scaled to be between zero and one to allow for each dataset to be plotted simultaneously while they may cover different timespans.  

\begin{figure}[htbp]
\centerline{\includegraphics[width = \columnwidth]{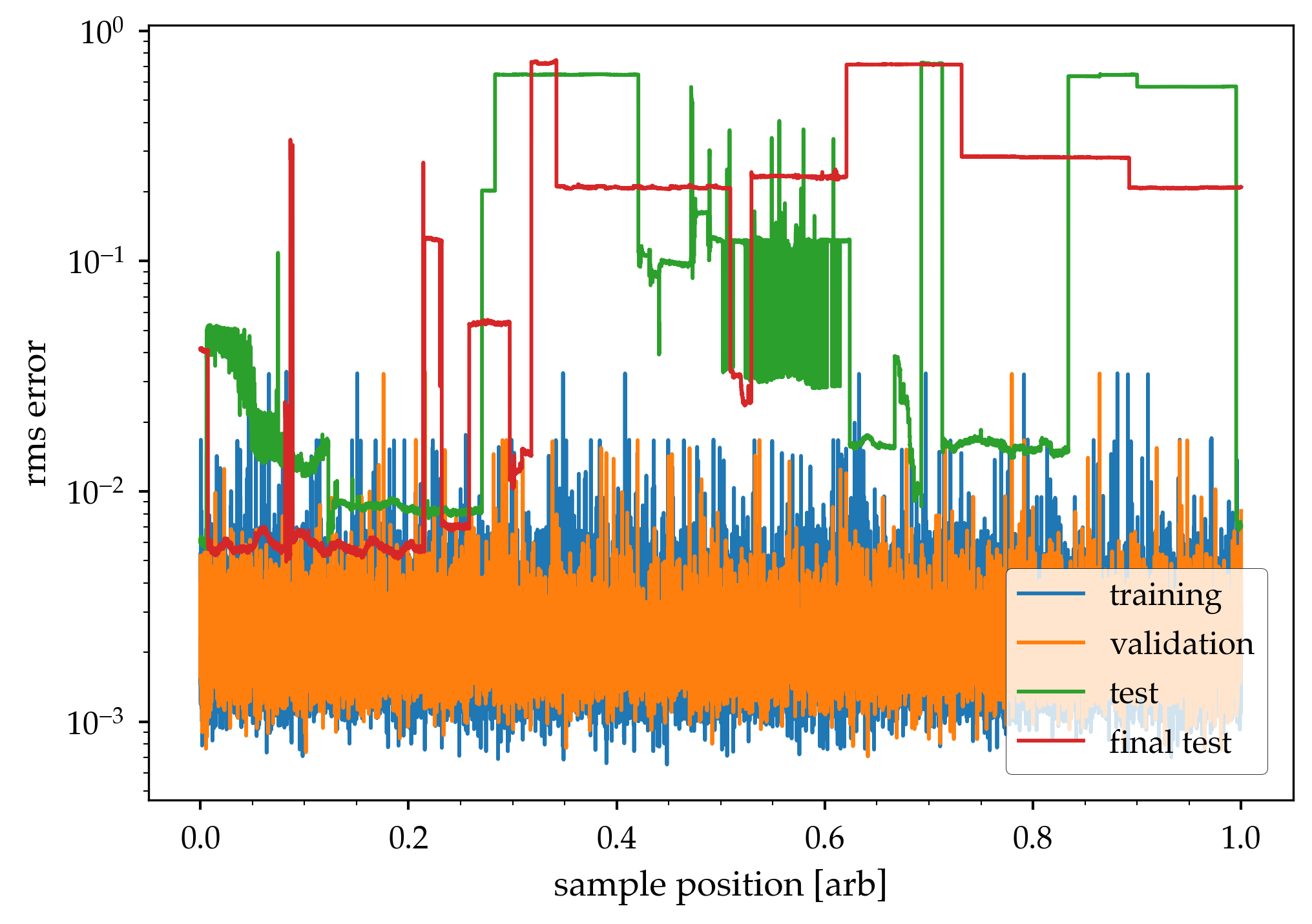}}
\caption{RMS error as a function of sample position for the reference data (training and validation) and fault data (test and final test) . The green dataset is used later for tuning detection thresholds in the semisupervised case while the red data are held aside for final testing.}
\label{fig:7}
\end{figure}

Figure \ref{fig:7} shows the RMS reconstruction error as a function of sample number for the reference data and the fault data. As with Figure \ref{fig:6}, the RMS reconstruction error on the fault data is on average two orders of magnitude larger than the reference data. This indicates that the machine is in a fundamentally different state leading up to the magnet faults. Our autoencoder is able to detect this state change using either of the two evaluation metrics discussed by being trained only on data during normal operation. 

\section{Region of convergence and semisupervised evaluation}
We determine the threshold for categorizing the anomalies using two different methods. The first method is entirely unsupervised, and the autoencoder assumes none of the reference data should be flagged as anomalous. The second method is semisupervised, and determines the threshold for anomalies based upon the performance of the autoencoder on the fault data. In this way, the semisupervised method aims to maximize the number of true positives while minimizing false positives. Half of the fault data runs were used for tuning the threshold while the other half were held aside for final testing to ensure robustness of our threshold choice.

We define a false positive as any datapoint in the reference set that is flagged as anomalous. Conversely, a true positive is defined as a sequence within the fault data that contains at least a single point that is flagged as anomalous. This definition permits us to optimize the threshold for selection such that true positives are maximized while maintaining low rates of false positives. 

The fully unsupervised routine, using the squared reconstruction error, flags 12\% of the fault data as anomalous while correctly identifying 17 of 26 anomalous runs. Using the RMS error metric, the unsupervised autoencoder flags 65 \% of the fault data as anomalous with 19 of 26 of the runs correctly identified as containing anomalous datapoints. In the unsupervised case we tested on all fault data.  

\begin{figure}[htbp]
\centerline{\includegraphics[width = \columnwidth]{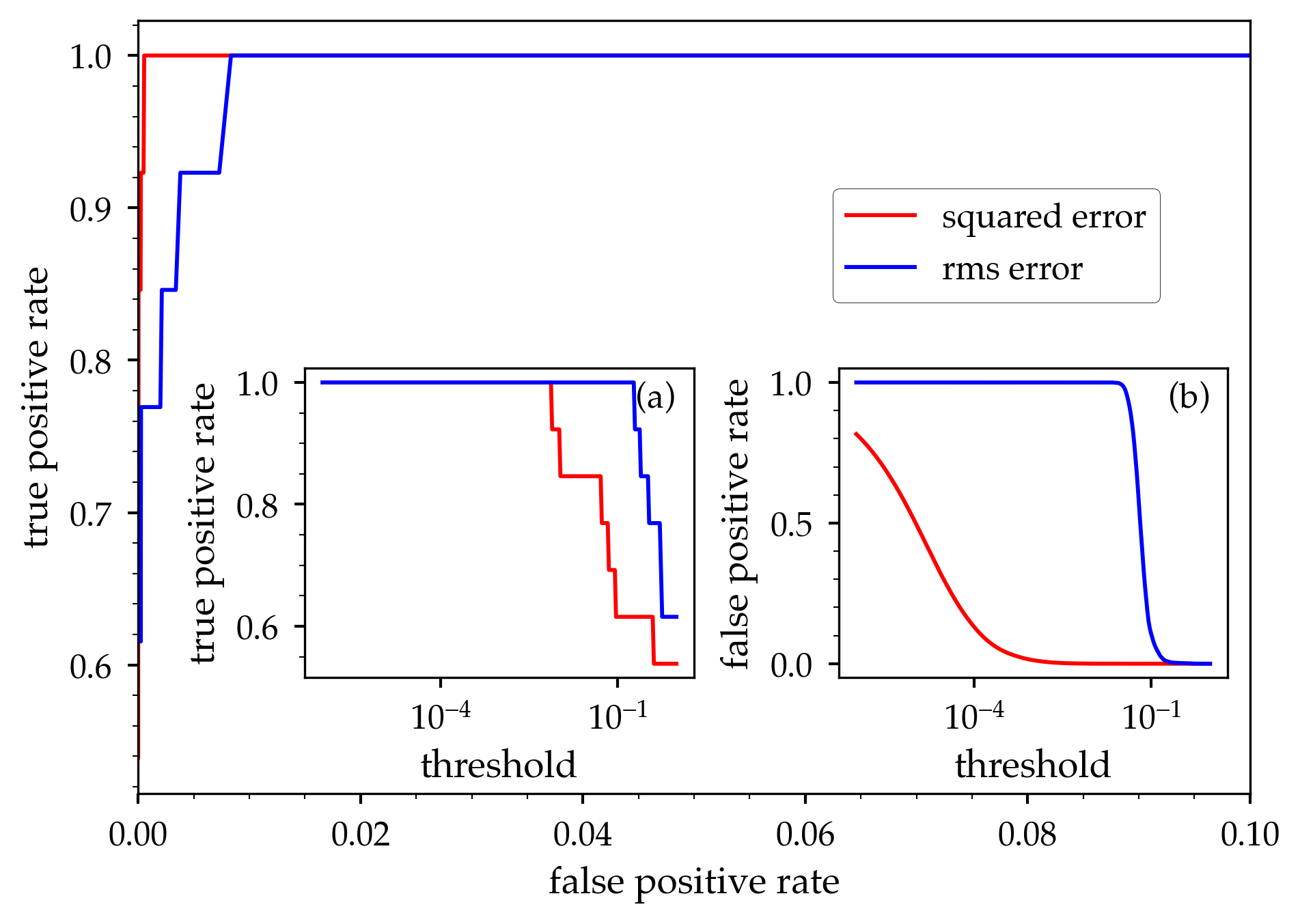}}
\caption{Region of convergence plot for the RMS error and squared error evaluation metrics. The main plot shows the true positive rate vs the false positive rate as a function of anomaly threshold. Inset a) shows the true positive rate as a function of the error threshold and inset b) shows the false positive rate as function of the error threshold. Note that the threshold is normalized to the peak value of the reconstruction error computed on the reference data.}
\label{fig:8}
\end{figure}

For the semisupervised case, we vary the reconstruction threshold in order to optimize the number of true positives while minimizing the number of false positives. Figure \ref{fig:8} shows the region of convergence (ROC) plot for both the squared error metric and the RMS error metric. More rapid convergence is obtained for the squared error metric than for the RMS metric.

\section{Fault detection and forecasting } 
\label{sec:fault}

\begin{figure}[htbp]
\centerline{\includegraphics[width = \columnwidth]{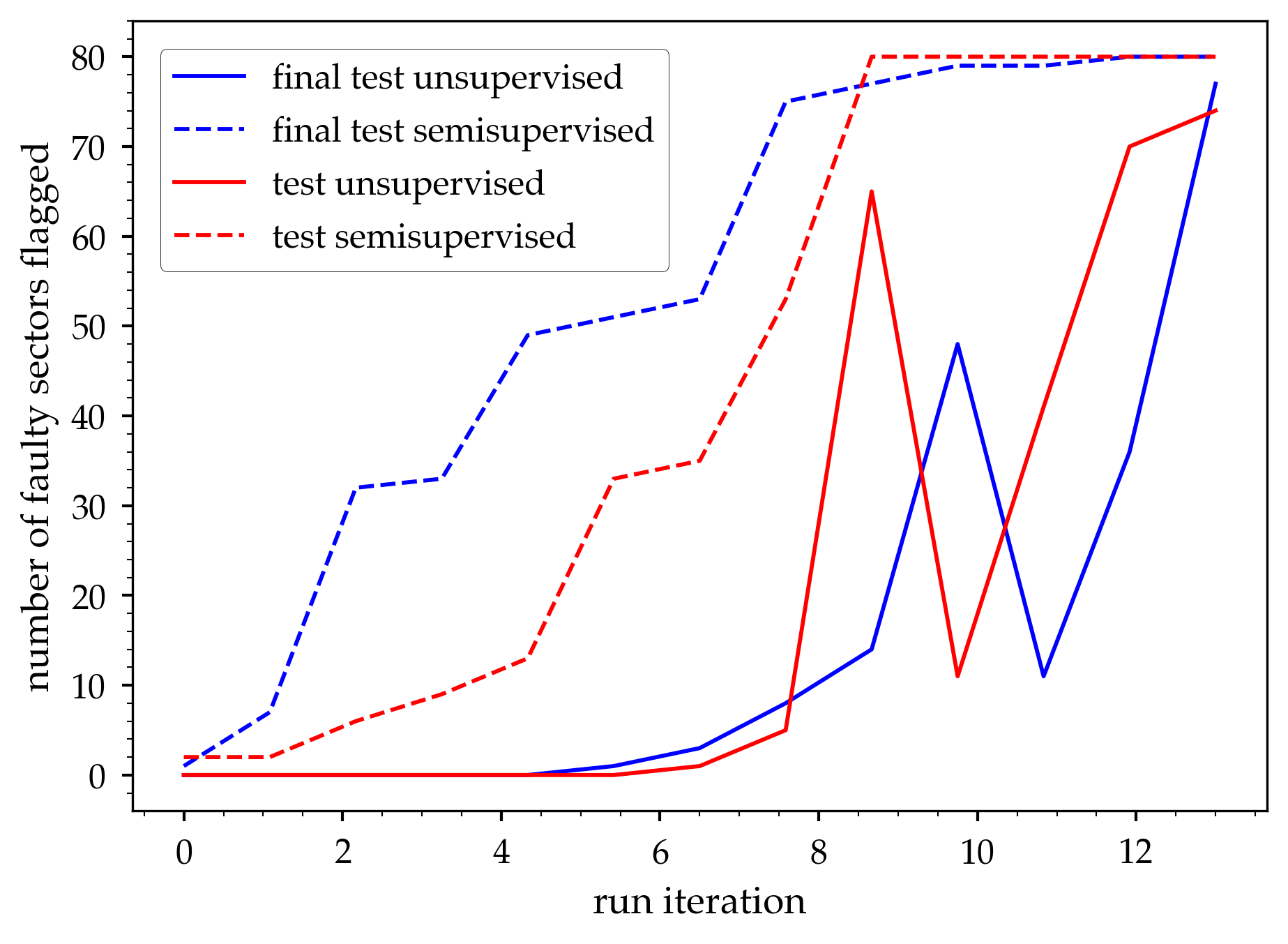}}
\caption{Number of faulty sectors for a given fault run. The data are sorted by the number of faulty sectors identified in the semisupervised case.}
\label{fig:9}
\end{figure}

Our previous analysis used the aggregate sector data to identify an anomalous machine state, with no indication of where the anomaly took place. We next examine the performance of the autoencoder in identifying sector-specific faults. As with our previous efforts, we consider both unsupervised and semisupervised approaches. To quantify our results, we make use of the fact that for each run in the test dataset, only a single sector experiences a magnet failure. Therefore, if the autoencoder identifies more than one sector as anomalous then either more data or a different evaluation metric will be required to improve the accuracy of the prediction. Because the RMS error metric cannot distinguish between sectors, we only consider the squared error metric. Figure \ref{fig:9} shows the number of anomalous sectors identified using the squared error metric and sorted for semisupervised case. As can be seen, multiple sectors are flagged across both scenarios, indicating a failure of the model to accurately identify the faulty sector, although the unsupervised routine appears to perform better. This suggests that additional information is necessary to accurately forecast specific magnet failures from these data. 

\begin{figure}[htbp]
\centerline{\includegraphics[width = \columnwidth]{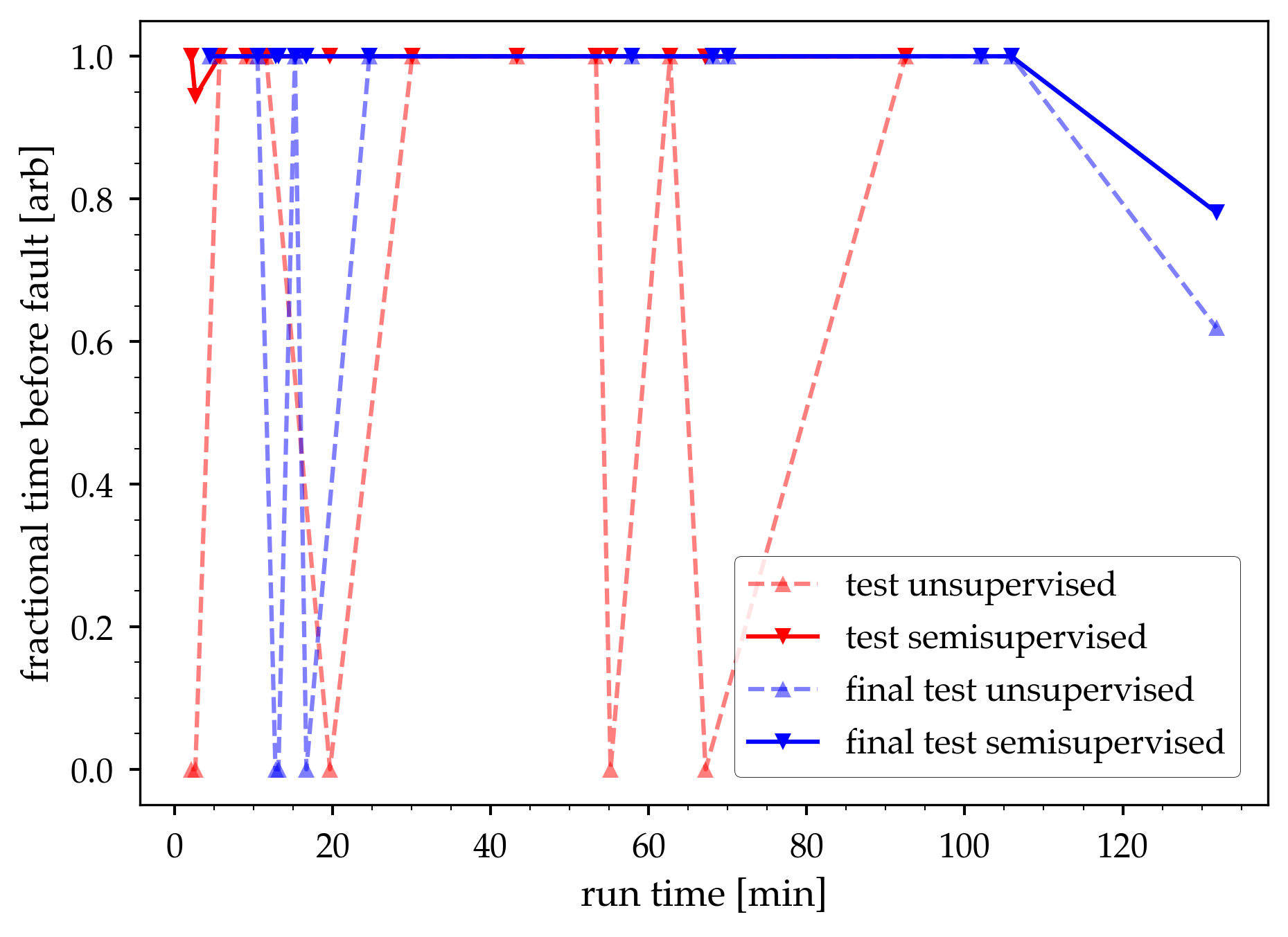}}
\caption{First indication of an anomaly as a function of the run time for the fault data using the RMS error metric. Red is the data used to tune the detection threshold while blue is the final test data that is not used in any of the training or parameter tuning. The dashed lines represent the unsupervised case while the solid line is the semisupervised case. }
\label{fig:11}
\end{figure}

\begin{figure}[htbp]
\centerline{\includegraphics[width = \columnwidth]{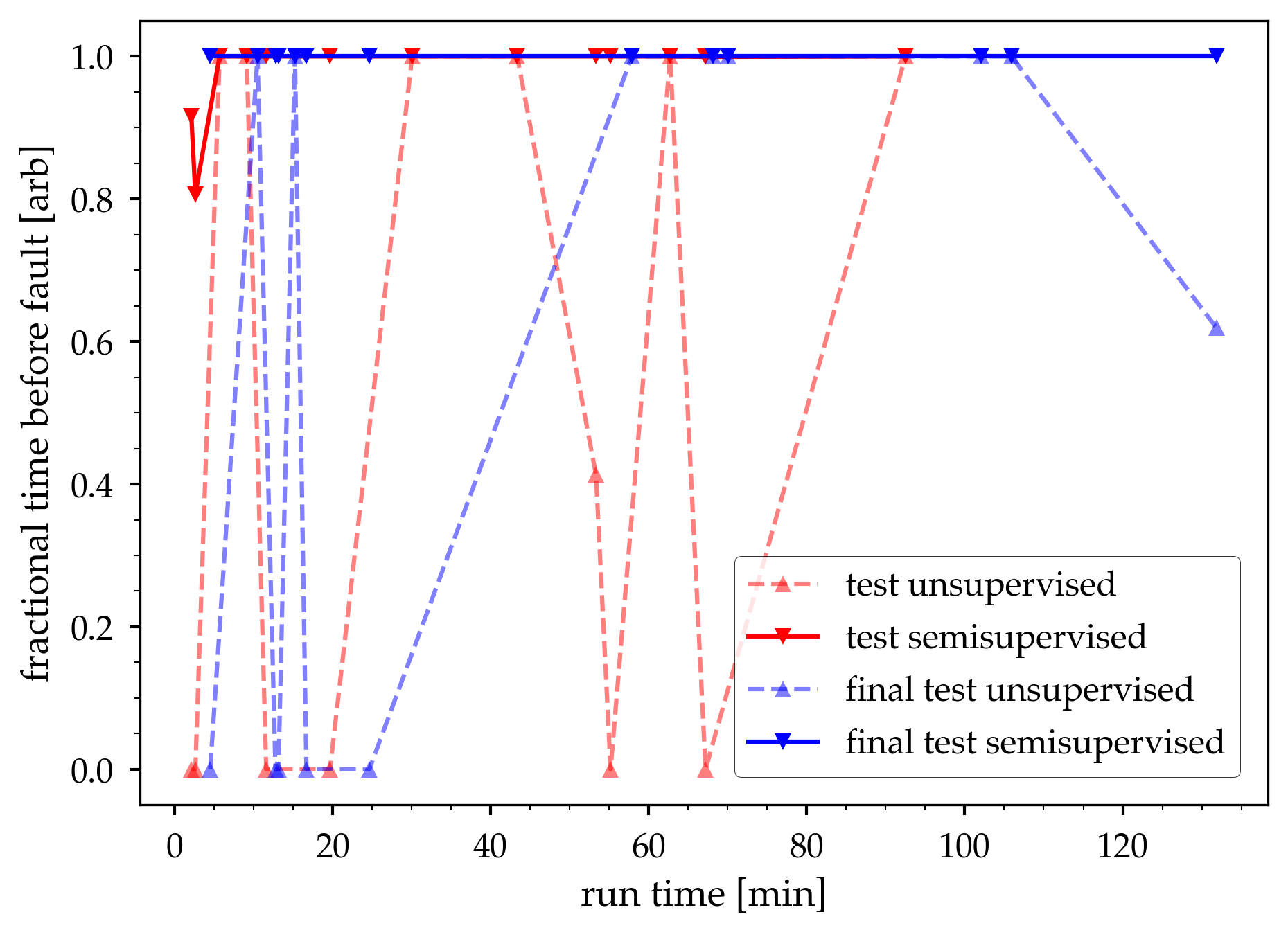}}
\caption{First indication of an anomaly as a function of the run time for the fault data using the squared error metric. Red is the data used to tune the detection threshold while blue is the final test data that is not used in any of the training or parameter tuning. The dashed lines represent the unsupervised case while the solid line is the semisupervised case}
\label{fig:12}
\end{figure}

However, the identification of precursors is much more promising. Figures \ref{fig:11} and \ref{fig:12} show analysis plots that compare unsupervised learning with semisupervised learning for the identification of precursors that result in magnet failures. Dashed lines depict the results from unsupervised learning and the solid lines for semisupervised learning. Because the fault runs are not the same length, the prediction is normalized to the total time of that run. For the RMS error metric, Figure \ref{fig:11} shows that precursors can be easily identified by the autoencoder for a large fraction of the runs; in some cases anomalous behavior is identified hours before the fault occurs. When applying the semisupervised threshold, the vast majority of the runs are correctly identified as anomalous at the start of the run, independent of run length. 

Figure \ref{fig:12} shows the same data as Figure \ref{fig:11} but for the squared error metric. Here we also see that many of the runs have precursors that are detectable long before the fault occurs. Moreover when we use the semisupervised thresholds, all but one of the runs has the majority of the data in a fault condition. This shows that the autoencoder is well suited to identify anomalous states in the machine and that precursors to the faults are detectable hours before the fault occurs even using aggregate parameters for a small subset of the dataset. 

\section{Conclusions} 
We have demonstrated the use of autoencoders to detect precursors to faults in the APS storage ring using both unsupervised and semisupervised learning. Using autoencoders we can reliably detect a change in the machine state independent of time before the fault occurs for our dataset. It is likely that the machine state change is detectable much sooner than our longest available dataset of just over 120 minutes. Analysis of more data over a wider range of operating configurations is necessary to test this hypothesis. 

As part of this work we also studied using supervised learning on a sequence of data to predict whether a fault will occur and the expected time to the fault. While the network was easily able to learn the binary output of whether or not the fault would occur, our models were unsuccessful in accurately predicting the timing of the fault. This is likely due to the fact that the test data largely samples machine states that were already heading towards a fault, without providing measurements during the earlier transition towards a faulty state. This conclusion is consistent with the data presented in Figures \ref{fig:11} and \ref{fig:12}, which show that the machine is in a different state during virtually all of the fault data. As a result, the data does not appear to provide the necessary indicators of the change from normal to faulty state, and subsequently cannot capture the total time-to-fault. Thus, while the autoencoder can clearly differentiate between the two machine states, more data are needed to take the next step in predicting when a fault will occur. 

We conclude that a) one can correctly identify a non-standard or anomalous machine state without directly sampling it (i.e. using unsupervised learning) and b) these anomalous states are detectable hours before a fault occurs. The ability to reliably distinguish between normal and anomalous states is critical to the operations of modern accelerators. Our results show that autoencoders are a promising tool for this application.

\section*{Acknowledgment}

The authors wish to thank Dr. Michael Borland and the Advanced Photon Source for contributing the data that allowed us to complete this study. This material is based upon work supported by the U.S. Department of Energy, Office of Science, Office of Nuclear Physics under Award Number DE-SC0019682.

\bibliographystyle{IEEEtran}
\bibliography{IEEEabrv,anomaly_detection}

\end{document}